\documentclass[a4paper,11pt]{article}
\pdfoutput=1

\usepackage{jcappub} 
\usepackage{aas_macros} 

\newcommand{\lsim}{\protect\raisebox{-0.8ex}{$\:\stackrel{\textstyle <}{\sim}\:$}} 
\newcommand{\gsim}{\protect\raisebox{-0.8ex}{$\:\stackrel{\textstyle >}{\sim}\:$}}

\newcommand{\eV}{\ensuremath{\ {\rm eV}}}

\title{Forecasts of cosmological constraints from Type Ia supernovae including the weak-lensing convergence}

\author[a]{Ryuichiro Hada}

\author[b]{and Toshifumi Futamase}

\affiliation[a]{Kavli IPMU (WPI), UTIAS, The University of Tokyo, Kashiwa, Chiba 277-8583, Japan}

\affiliation[b]{Department of Astrophysics and Atmospheric Sciencesz, Kyoto Sangyo University, Kyoto, Kyoto 603-8555, Japan}

\emailAdd{ryuichiro.hada@ipmu.jp}
\emailAdd{tof@cc.kyoto-su.ac.jp}

\abstract{
We investigate how the cosmological constraints from SNe Ia are improved by including the effects of weak-lensing convergence. To do so, we introduce the lognormal function as the convergence PDF modeling the lensing scatter of SN Ia magnitude, and apply a sample selection for SNe Ia to avoid strongly lensed samples. Comparing with the contribution of other uncertainties (e.g., the intrinsic magnitude scatter), we find that the lensing effect is dominant at $z > 1$. Then forecasting the parameter constraints for the {\it Wide-Field InfraRed Survey Telescope} survey, we show that considering the weak-lensing effect, the constraints on the density parameters $\Omega_m$ or $\Omega_{\Lambda}$, and the dark energy equation of state $w$ are improved, especially for SNe Ia samples at higher redshift $z > 1$. Furthermore, we see that the degeneracy between the total mass of neutrino $\Sigma m_{\nu}$ and the (cold) dark matter density parameter $\Omega_{c}$ can be resolved.
}

\begin{document}
\maketitle
\flushbottom

\section{\label{sec1}Introduction}

Type Ia supernovae (SNe Ia), which are widely known as cosmological standard candles, play an important role in measuring the cosmological distance. In particular, two independent research groups measured the distance-redshift relation for high-redshift SNe Ia and found that our universe is expanding at an accelerating rate~\cite{1998AJ....116.1009R,1999ApJ...517..565P}, which suggests that our universe is filled with dark energy. Since then, there have been a large number of SN surveys over a wide redshift range (see, e.g., \cite[][]{2018ApJ...859..101S}, for recent results), which allows us to extend the distance-redshift relation up to $z=2$ or higher. Furthermore, it is expected that we can obtain a larger number of SNe Ia at higher redshift in the ongoing and upcoming SN surveys: the Dark Energy Survey (DES)~\cite{2016MNRAS.460.1270D, 2012ApJ...753..152B}, the {\it Wide-Field InfraRed Survey Telescope} ({\it WFIRST})~\cite[][]{2015arXiv150303757S} and  the Large Synoptic Survey Telescope (LSST)~\cite[][]{2008arXiv0805.2366I,2009arXiv0912.0201L}, etc.

However, light rays from SNe Ia at higher redshift pass through the large scale structure (LSS) for a longer time, and the observed magnitudes are more likely to be (de)magnified due to the effects of weak-lensing convergence~\cite{2010MNRAS.405..535J,2010A&A...514A..44K,2014ApJ...780...24S}. It means that we need to take account the weak-lensing effect to correctly handle with the magnitude scatter of SNe Ia at high redshift. While the intrinsic magnitude scatter or the st	atistical measurement and model uncertainties (due to light-curve fitting or redshift measurement) does not depend on redshift, it was found that the lensing scatter is approximately proportional to the SN Ia redshift: $\sigma_{\rm lens} = 0.055z$ \cite{2010MNRAS.405..535J}, which has been used in past SN surveys~\cite{2011ApJS..192....1C, 2018ApJ...859..101S}. In this expression, the lensing scatter looks a type of systematic uncertainty, however, it definitely includes the LSS information through the weak-lensing convergence in addition to the composition in a homogeneous and isotropic universe. We are therefore able to extract that by modeling the probability distribution function (PDF) of the SNe Ia magnitude for a given cosmology.

The magnitude PDF of SNe Ia, i.e., the convergence PDF, has been widely studied in the last 20 years (see, e.g., \cite[][]{2013PhRvD..88f3004M,2016ApJ...828..112H}, for summary). In particularly, it is found that the convergence PDF is also well described by the lognormal function~\cite{2002ApJ...571..638T,2006ApJ...645....1D,2011ApJ...742...15T}, which reflects the fact that the PDF of the matter density field is well modeled by the lognormal function~\cite{2001ApJ...561...22K}. We can calculate the lognormal PDF from the variance of the lensing convergence, which is described as an integral of the matter power spectrum (see Sec.~\ref{conv} for details), and then discuss about the dependence of the convergence PDF on the cosmological parameters characterizing the LSS. For instance, while the clustering amplitude $\sigma_{8}$ describes how bumpy the matter density field is, it can not be constrained at all by the mean value of the SN Ia magnitudes at each redshift. However, modeling the scatter of the SN Ia magnitudes due to the lensing effect allows us to constrain on $\sigma_{8}$~\cite{1999ApJ...519L...1M,2014PhRvD..89b3009Q,2017MNRAS.465.2862S}. 

Furthermore, the total mass of neutrinos $\Sigma m_{\nu}$ have effects on the LSS because massive neutrinos slow down the growth of matter density perturbations on small scales (e.g., \cite{2006PhR...429..307L}). In our previous paper~\cite[][]{2016ApJ...828..112H}, we have estimated how the total mass of neutrinos $\Sigma m_{\nu}$ is constrained from the lensing scatter of SNe Ia and found that marginalizing $\Sigma m_{\nu}$ and the magnitude error other than lensing, the constrain expected from both {\it WFIRST} and LSST (main) is $\Sigma m_{\nu} < 1.1 \eV $(95\% CL). While the current upper bound $\Sigma m_{\nu} \lsim 0.15 \eV $(95\% CL)~\cite{2017MNRAS.470.2617A,2017PhRvD..96l3503V} was set from the galaxy clustering data set of the Baryon Oscillation Spectroscopic Survey (BOSS) combined with the Planck cosmic microwave background (CMB) measurements, etc., constraints on neutrino masses from the Ly$\alpha$ forest has been also discussed (e.g.,~\cite{2015JCAP...11..011P}). The lensing convergence of SNe Ia are not suffered from the issue of galaxy bias unlike in the case of galaxy clustering and the Ly$\alpha$ forest, and then expected to be an independent probe of neutrino masses.\footnote{The comparison with another gravitational lensing effect reflecting the LSS, cosmic shear, was discussed in Sec.~2.2 of \cite[][]{2016ApJ...828..112H}. \label{foot_1}}

Then, in this paper, we particularly focus on how the cosmological constraints from SNe Ia are improved by considering the lensing effect and forecast constraints on not only the total mass of neutrinos $\Sigma m_{\nu}$ but also the density parameter $\Omega_m$ and $\Omega_{\Lambda}$, or the dark energy equation of state $w$. Following our previous work~\cite{2014JCAP...12..042H, 2016ApJ...828..112H}, we apply a sample selection for SNe Ia to avoid complications associated with small-scale structures (e.g., strong lensing) and set a critical value of the wave number in an integral of the matter power spectrum. Furthermore, we use the lognormal function as the convergence PDF to more realistically model the distribution of SN Ia magnitude while a Gaussian PDF was approximately used in the previous work.

The paper is organized as follows. In Sec.\ref{PDF}, we first introduce a convergence PDF and a sample selection of SNe Ia to exclude some difficulties caused by small-scale structures, in order to obtain the PDF for the lensing (de)magnification. We then convolute the lensing PDF and the PDF for the magnitude scatter other than lensing, and gain the total PDF of the SN Ia magnitude. In Sec.~\ref{Fore}, we forecast constraints on some cosmological parameters using the Fisher information matrix for the {\it WFIRST} survey, and compare the results with and without the lensing effect. Finally, Sec.~\ref{conc} is devoted to discussion and conclusions.

\section{\label{PDF}PDF for the apparent magnitude of SNe Ia}

In this section, we show the PDF for the apparent magnitude of observed SNe Ia. Using the relation between the observed apparent magnitude $m$ and the amplitude of energy flux $f$ for a light source,
\begin{eqnarray}
	m = - 2.5 \log_{10} f + {\rm const}, \label{eq:m-f}
\end{eqnarray}
we can write the difference between the observed apparent magnitude $m_{\rm obs}$ and the true one $m_{\rm true}$ in terms of the lensing effect $x_{\rm lens}$ and other effects $x_{\rm othe}$:
\begin{eqnarray}
	\delta m 
		&=& m_{\rm obs} - m_{\rm true} 
		\nonumber \\
		&=& x_{\rm lens} + x_{\rm othe}
		\nonumber \\
 		&=& - 2.5 \log_{10} \frac{f_{\rm lens}}{f_{\rm no \mathchar`- lens}} + x_{\rm othe},
	\label{eq:dm}
\end{eqnarray}
where $f_{\rm lens}$ and $f_{\rm no \mathchar`- lens}$ are the amplitude of energy flux magnified (or demagnified) by the lensing effect and the one not magnified (or demagnified), respectively. 

In the following, we introduce a convergence PDF and provide the PDF of $x_{\rm lens}$ in Sec.~\ref{conv} and consider a selection for SNe Ia and the corresponding critical value of the wave number to overcome some difficulties caused by small-scale structures in Sec.~\ref{smal}. Furthermore, in Sec.~\ref{othe}, we consider how to deal with other uncertainties $x_{\rm othe}$ (e.g., the intrinsic scatter in SNe Ia luminosities).

\subsection{\label{lens}Lensing magnification}

The flux magnification in gravitational lensing, $\mu = f_{\rm lens}/f_{\rm no \mathchar`- lens}$, is described by the convergence $\kappa$ and sheer $\gamma$ (e.g., \cite{2001PhR...340..291B}) as follows,
\begin{eqnarray}
	\mu
 		&=& \frac{1}{(1-\kappa)^2 - \gamma^2}.
	\label{eq:mu}
\end{eqnarray}
This equation means that if we try to estimate the magnification of light rays exactly, we need both the convergence and shear. However, \cite{1989PhRvD..40.2502F} found that the contribution of the shear to the net magnification is small enough compared with the convergence unless light rays are strongly magnified (or demagnified) by structures below galactic scales (note that we discuss about the exclusion of such samples in Sec~\ref{smal}). In fact, high-resolution ray-tracing simulations~\cite{2011ApJ...742...15T} indicate that even when neglecting the shear term, Eq.~(\ref{eq:mu}) successfully describes the relation between the magnification and convergence. Therefore, combining Eqs.~(\ref{eq:dm}) and (\ref{eq:mu}), the contribution of the lensing effect in $\delta m$ can be described by
\begin{eqnarray}
	x_{\rm lens}
 		&=& - 2.5 \log_{10} \mu
		\nonumber \\
 		&\simeq& 5 \log_{10} |1- \kappa|. 
	\label{eq:x_lens}
\end{eqnarray}
We note that the approximation of neglecting the shear term breaks down as the convergence gets larger (see Fig.~8 in \cite{2011ApJ...742...15T}).

\subsubsection{\label{conv}Convergence PDF}

The convergence of the bundle of light rays from a source at $z=z_{s}$ is represented as an integral over the matter density fluctuation along the line of sight (e.g., \cite{2001PhR...340..291B}):
\begin{eqnarray}
	\kappa(z_s,\hat{\bf{n}})
		&=& \frac{3H_0^2\Omega_{m}}{2} \int_{0}^{\chi_{s}} d\chi\frac{r(\chi)r(\chi_s-\chi)}{r(\chi_s)}(1+z)\delta_m(z,\hat{\bf{n}}), \label{eq:K-z}
\end{eqnarray}
where $\chi(z)$ is the comoving distance ($\chi_s \equiv \chi(z_s)$), $r(\chi)/(1+z)$ is the angular-diameter distance, and $\hat{\bf{n}}$ is the source direction. In addition, $H_{0}$ is the present Hubble parameter, $\Omega_{m}$ is the present matter density parameter, and $\delta_m$ is the relative perturbation of matter. 

Here we are interested in the PDF of the convergence $\kappa$. While we can see from the equation above that the convergence PDF is determined by the PDF of the density fluctuation along the line of sight, we can actually calculate the convergence PDF under a hierarchical ansatz for the higher order moments of $\delta_m$ and an approximation holding for the limit of $z_s \to 0$~\cite{2000MNRAS.318..109M, 2000A&A...356..771V}. An analysis using $N$-body simulations~\cite{2001ApJ...561...22K} showed that the PDF of the matter density field is well described by the lognormal distribution and thus we obtain the following lognormal function as the convergence PDF:
\begin{eqnarray}
	\frac{dP}{d\kappa} [\kappa]
		&=& \frac{1}{\sqrt{2\pi}\sigma_{\rm ln}}\exp\left[ - \frac{\{ \ln (1 + \kappa / | \kappa_{\rm min} |) + \sigma^2_{\rm ln} /2 \}^2}{2 \sigma^2_{\rm ln}} \right] \frac{1}{\kappa + | \kappa_{\rm min} |}. \label{eq:P_K}
\end{eqnarray}
Here $\kappa_{\rm min}$ corresponds to the minimum value when light rays pass through the empty region all the way:
\begin{eqnarray}
	\kappa_{\rm min}(z_s)
		&=& -\frac{3H_0^2\Omega_{m}}{2} \int_{0}^{\chi_{s}} d\chi\frac{r(\chi)r(\chi_s-\chi)}{r(\chi_s)}(1+z), \label{eq:K_min}
\end{eqnarray}
and $\sigma_{\rm ln}$ is defined by 
\begin{eqnarray}
	\sigma^2_{\rm ln} \equiv \ln \left( 1 + \frac{\langle \kappa^2 \rangle}{ | \kappa_{\rm min} | ^2}\right), \label{eq:sigma_ln}
\end{eqnarray}
where $\langle \kappa^2 \rangle$ is the variance of convergence: 
\begin{eqnarray}
	\langle \kappa^2 (z_s) \rangle
		&=& \left(\frac{3H_0^2\Omega_{m}}{2}\right)^2 \int_{0}^{\chi_{s}} d\chi \left[\frac{r(\chi)r(\chi_s-\chi)}{r(\chi_s)} (1+z)\right]^2
		\nonumber \\
		&& \times \int_0^\infty \frac{d\ln k}{2\pi}k^2 P_{\rm nl}(z,k),  \label{eq:cov_K}
\end{eqnarray}  
where $z = z(\chi)$ and $P_{\rm nl}(z,k)$ is the nonlinear matter power spectrum. It was actually found, in \cite{2002ApJ...571..638T}, that the result of ray-tracing simulations was well modeled by the lognormal function Eq.~(\ref{eq:P_K}). 

Using the relation between $x_{\rm lens}$ and $\kappa$, Eq.~(\ref{eq:x_lens}), we can compute the PDF of $x_{\rm lens}$ from the convergence PDF, Eq.~(\ref{eq:P_K}):
\begin{eqnarray}
	L_{\rm lens}[x_{\rm lens}] \equiv \frac{dP}{dx_{\rm lens}} [x_{\rm lens}]
		&=& \left| \frac{d\kappa}{dx_{\rm lens}} \right|\frac{dP}{d\kappa} [\kappa (x_{\rm lens})]
		\nonumber \\
		&=& \frac{\ln 10}{5}\left\{1- \kappa(x_{\rm lens})\right \} \frac{dP}{d\kappa} [\kappa (x_{\rm lens})], \label{eq:P_x_lens}
\end{eqnarray}
where
\begin{eqnarray}
	\kappa (x_{\rm lens})
  		&=& 1 - \exp \left[ \frac{\ln 10}{5} x_{\rm lens} \right]. 
	\label{eq:kappa}
\end{eqnarray}
Here we used the fact $\kappa < 1$.

\subsubsection{\label{smal}Small-scale structures}

Compared with cosmic shear observation, the convergences of light rays from SNe are influenced by smaller-scale structures because SNe are point sources, which means that we can obtain the information on smaller scales. However, some of the observed SNe Ia will be {\it strongly} magnified by a gravitational lens and such samples are usually not included in the final cosmological SNe Ia catalog. Then we need to carefully pay attention to how to handle with highly magnified samples.  

As we mentioned in Sec.~\ref{conv}, we obtain the convergence PDF in the same lognormal function as the PDF of the density fluctuation, applying an approximation holding for the limit of $z_s \to 0$. In the limit of $z_s \to 0$, the effect of the integral along the line of sight in Eq.~(\ref{eq:K-z}) is able to be safely neglected and we can understand why the net convergence PDF along the line of sight can be described well by the same function as the density PDF at a specific redshift. However, the higher the source redshift $z_s$ becomes, the larger the effect of the integral along the line of sight is. Therefore it is hard that the lognormal function describes the convergence PDF away from $\kappa = 0$, especially the high magnification tail. Some ray-shooting simulations actually found that the (modified) lognormal distribution fails to model the high convergence tail~\cite{2002ApJ...571..638T, 2011ApJ...742...15T}. 

In our previous work~\cite{2016ApJ...828..112H}, in order to systematically exclude the strongly lensed samples, we introduced a critical radius $\theta_{c}$ and considered the following sample selection for SNe Ia: “we only use such SNe Ia that the centers of the foreground galaxies are not included in the area within the critical radius $\theta_{c}$ of the SNe Ia.” Assuming that lensing objects are only galaxies with the singular isothermal sphere (SIS) mass profile and performing order estimation, we found that it is required that $\theta_{c}$ is larger than the order of $1-10 ''$ and the light rays for SNe Ia selected with such a critical radius are not affected by structures with the smaller masses than the corresponding critical mass, $M_{\rm c} \simeq 10^{11} - 10^{12} M_{\odot}$.\footnote{In Sec.~2.2 of \cite{2016ApJ...828..112H}, we discussed about the necessity of excluding the highly magnified samples as well in the context of safely neglecting the shear term in Eq.~(\ref{eq:mu}), and explained why the contribution of clusters as lensing objects can be ignored in the sample selection. \label{foot_2}}

Accordingly, we need to connect the correspondence between the sample selection and the critical mass to the expression of the convergence PDF in Sec.~\ref{conv}. To do so, we eliminate the effects from structures with a smaller scale than $M_{\rm c}$ using the Gaussian smoothing filter in Eq~(\ref{eq:cov_K}): 
\begin{eqnarray}
	\langle \kappa^2 (z_s) \rangle
		&=& \left(\frac{3H_0^2\Omega_{m}}{2}\right)^2 \int_{0}^{\chi_{s}} d\chi \left[\frac{r(\chi)r(\chi_s-\chi)}{r(\chi_s)}(1+z)\right]^2
		\nonumber \\
		&& \times \int_0^\infty \frac{d\ln k}{2\pi}k^2 P_{\rm nl}(z,k) \ \exp [-k^2/k_{c}^2(z)],  \label{eq:cov_K_cut}
\end{eqnarray}  
where $k_{c}(z)$ is the critical wave number corresponding to the critical mass $M_{\rm c}$, which is given in Sec.~2.3 of \cite{2016ApJ...828..112H}.

\subsection{\label{othe}Other uncertainties}

We next consider uncertainties causing the scatter in the observed apparent magnitude other than the lensing effect, $x_{\rm othe}$. These uncertainties can be mainly separated into the following two parts: (1) the distance uncertainty for each SN Ia, which includes both statistical measurement uncertainties (e.g., the uncertainty in measuring redshift) and model uncertainties (i.e., the uncertainty due to the light-curve fit); (2) the intrinsic scatter of SNe Ia as standard candles, which still remains after correcting SN Ia distances with the stretch and color~\cite{2014ApJ...780...24S,2017arXiv170201747H}. In this paper, we assume that the net uncertainty $x_{\rm othe}$ obeys the Gaussian distribution:
\begin{eqnarray}
	L_{\rm othe}[x_{\rm othe}] \equiv \frac{dP}{dx_{\rm othe}} [x_{\rm othe}]
		&=& \frac{1}{\sqrt{2\pi}\sigma_{\rm othe}}\exp\left[ - \frac{x_{\rm othe}^2}{2 \sigma^2_{\rm othe}} \right], \label{eq:P_x_othe}
\end{eqnarray}
where $\sigma_{\rm othe}^2$ is the variance of $x_{\rm othe}$.

Finally, convoluting the two PDFs: Eqs.~(\ref{eq:P_x_lens}) and (\ref{eq:P_x_othe}), we obtain the PDF for the total scatter in the observed apparent magnitude,   $x_{\rm tot}$ ($= x_{\rm lens} + x_{\rm othe}$):
\begin{eqnarray}
	L_{\rm tot}[x_{\rm tot}] = \int^{\infty}_{-\infty} dy \ L_{\rm lens}(x_{\rm tot} - y) \ L_{\rm othe}(y). \label{eq:P_x_tot}
\end{eqnarray}
Here $x_{\rm tot}$ is equivalent to $\delta m = m_{\rm obs} - m_{\rm true} $ in Eq.~(\ref{eq:dm}). Note that in \cite{2016ApJ...828..112H}, we assumed that the PDF of $x_{\rm lens}$, $L_{\rm lens}$, is given by the Gaussian distribution with the variance:\footnote{We also assumed that the convergence is small enough that the second or higher order can be neglected, and then used the following approximated relation: $x_{\rm lens} \simeq - (5 / \ln 10)\  \kappa$. In this point, the relation we use in this paper, Eq.~(\ref{eq:x_lens}), is valid for higher-convergence samples although we exclude the strongly magnified samples after all as discussed in Sec.~\ref{smal}. \label{foot_3}} $\sigma_{\rm lens}^2(z_s) = (5/\ln 10)^2 \langle \kappa^2 (z_s)\rangle$, and consequently used, as $L_{\rm tot}$, the Gaussian with the variance: $\sigma_{\rm tot}^2 = \sigma_{\rm lens}^2 + \sigma_{\rm othe}^2$.

\section{\label{Fore}Forecasts of cosmological constraints}

In this section, we estimate the extent to which cosmological constrains from  SNe Ia can be improved by taking account of the weak-lensing convergence due to LSS. To do so, in Sec.~\ref{detail}, we introduce a future survey, which we use in our forecast, and set some parameters. In Sec.~\ref{mag_PDF}, we show the PDF for the magnitude of SNe Ia that was introduced in Sec.~\ref{PDF}. We then consider the Fisher information matrix, and forecast constraints on some cosmological parameters in Sec.~\ref{fisher}.

\begin{table}[b]
  \begin{center}
	\caption{\label{table_1}SNe Ia per $\Delta z = 0.1$ bin in {\it WFIRST}}
    \begin{tabular}{c|l} \hline
      Redshift  &  \quad Number of SNe Ia \\ \hline 
   $z = 0.2$   & \quad  $0.6 \times 10^{2}$     \\
	  0.3     &  \quad $2.0 \times 10^{2}$      \\
	  0.4     &  \quad $4.0 \times 10^{2}$      \\
	  0.5     &  \quad $2.2 \times 10^{2}$      \\
	  0.6     &  \quad $3.2 \times 10^{2}$      \\
      0.7-1.7 &  \quad $1.4 \times 10^{2}$ (for each bin) \\ \hline
    \end{tabular}
	\vskip 3pt
  \end{center}
\end{table}

\subsection{\label{detail}Survey considered and parameter setting}

The {\it Wide-Field InfraRed Space Telescope} ({\it WFIRST};~\cite[][]{2015arXiv150303757S}) is a NASA 	mission in formulation with a planned launch in the mid-2020's. One of major {\it WFIRST} goals is to precisely constrain the nature of dark energy through multiple programs, including SNe Ia. In our forecast, we will use the expected numbers of SNe Ia from {\it WFIRST}, which are summarized in Table~\ref{table_1}. Note that we opt not to use SNe Ia at at $z<0.1$, where the Doppler terms could be dominant (see Sec.~2.2 in \cite[][]{2016ApJ...828..112H}). The {\it WFIRST}-AFTA 2015 Report~\cite{2015arXiv150303757S, 2017arXiv170201747H} assumes that the distance precision
per SN is $\sigma_{\rm meas}=$0.08[mag] and the intrinsic scatter of SNe is $\sigma_{\rm int}=$0.09[mag]. Considering the fact: $\sigma_{\rm othe}^2 = \sigma_{\rm meas}^2 + \sigma_{\rm int}^2$, we then use the following fiducial value for the variance $\sigma_{\rm othe}$, $\sigma_{{\rm othe},f}=0.12$.

Furthermore, following our previous paper~\cite[][]{2016ApJ...828..112H}, we use the same fiducial value of the critical mass, $M_{\rm c}$, deciding the critical wave number: $M_{{\rm c},f}=10^{11}M_{\odot}$. In the following analysis, we use {\sc camb}\footnote{\url{http://camb.info/} \label{foot_4}}~\cite{2000ApJ...538..473L} for calculating general utility function for cosmological calculations. In particular, we compute the nonlinear power spectrum in Eq.~(\ref{eq:cov_K}) through a modified {\sc halofit}~\cite{2003MNRAS.341.1311S,2012ApJ...761..152T,2012MNRAS.420.2551B}, which includes the effects of massive neutrinos and has been incorporated into {\sc camb}.

\begin{figure}[t]
\begin{tabular}{cc}
 \begin{minipage}{0.5\hsize}
  \begin{center}
   \includegraphics[width=80mm]{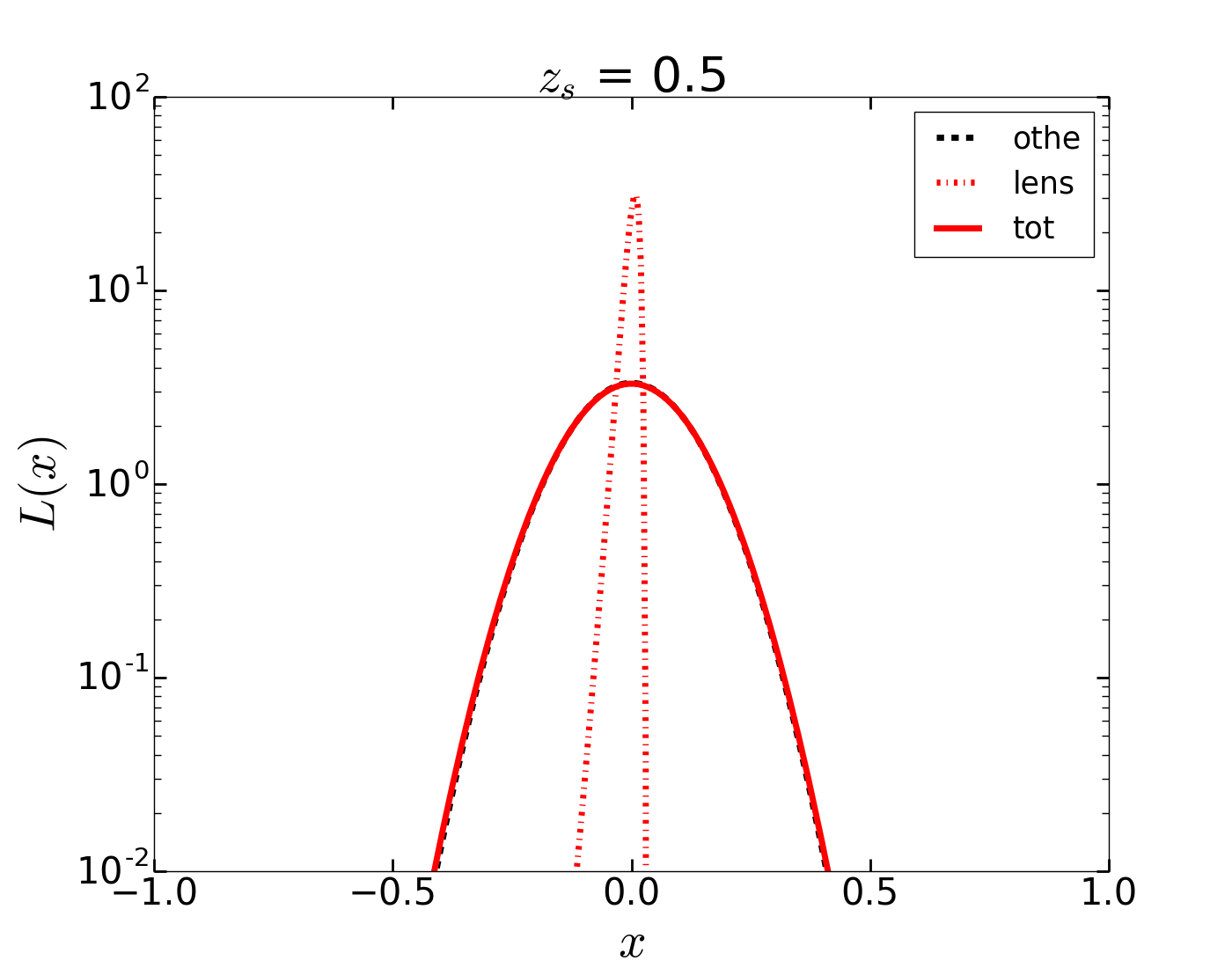}
  \end{center}
 \end{minipage}
 \begin{minipage}{0.5\hsize}
  \begin{center}
   \includegraphics[width=80mm]{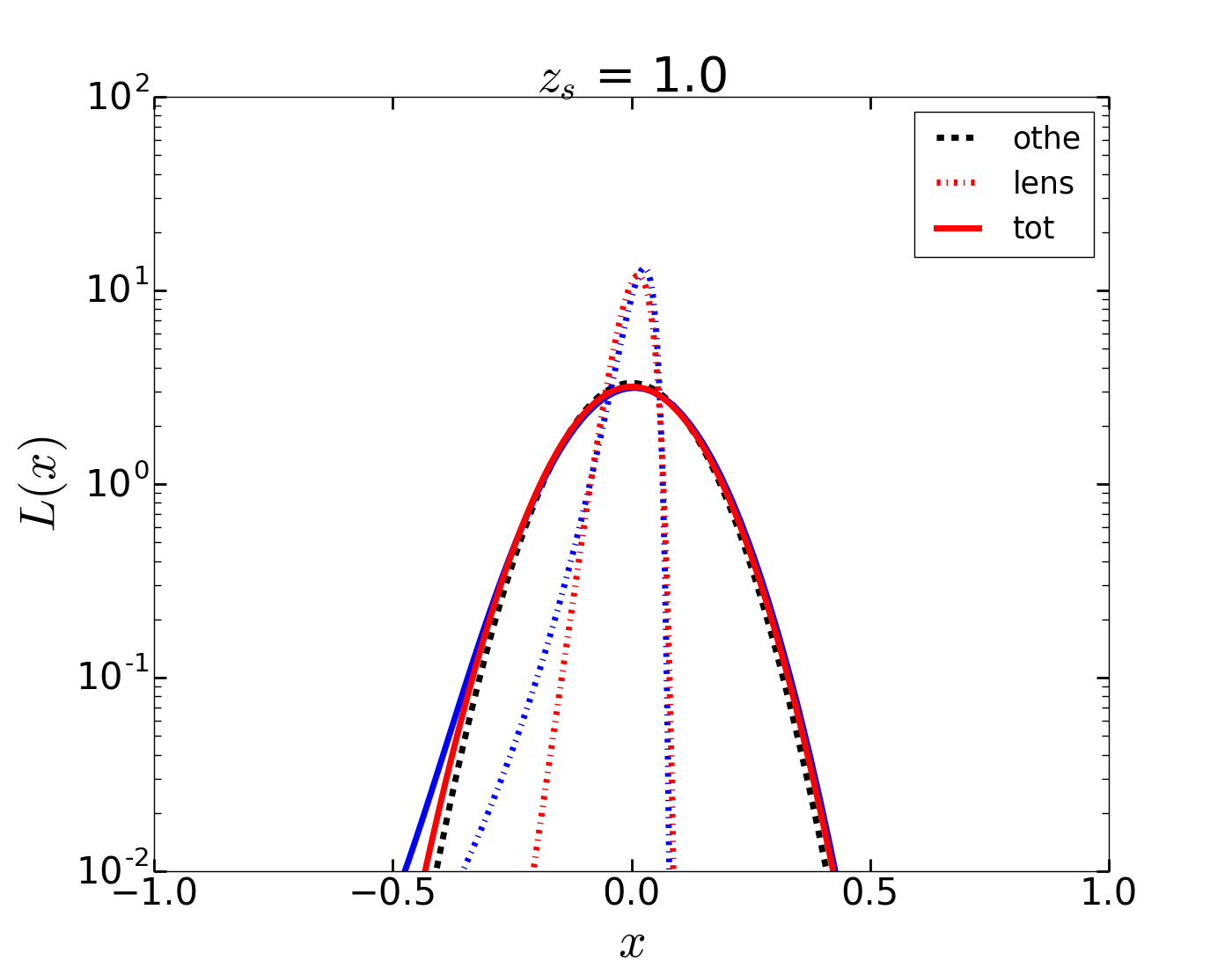}
  \end{center}
\end{minipage} \\
 \begin{minipage}{0.5\hsize}
  \begin{center}
   \includegraphics[width=80mm]{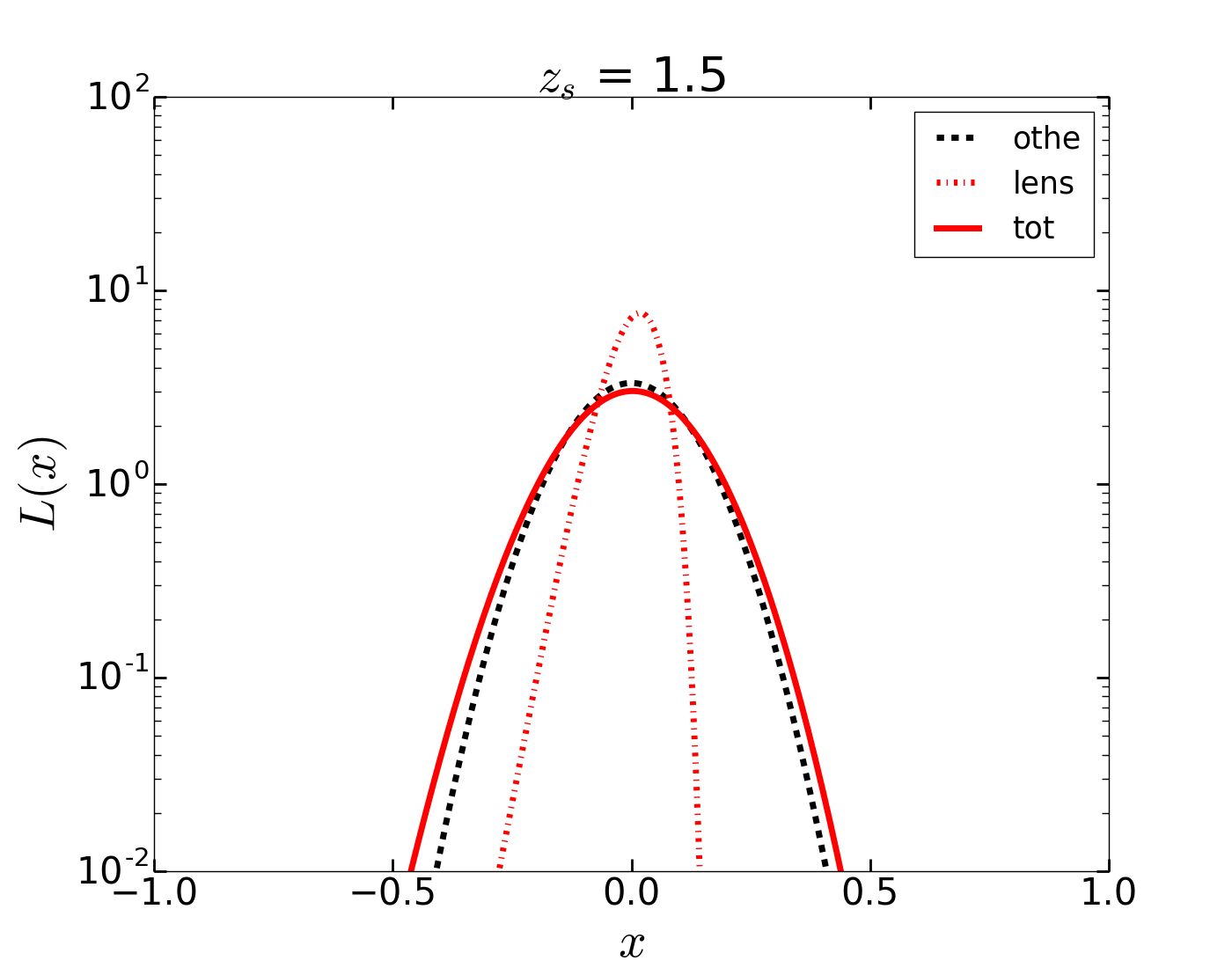}
  \end{center}
 \end{minipage}
 \begin{minipage}{0.5\hsize}
  \begin{center}
   \includegraphics[width=80mm]{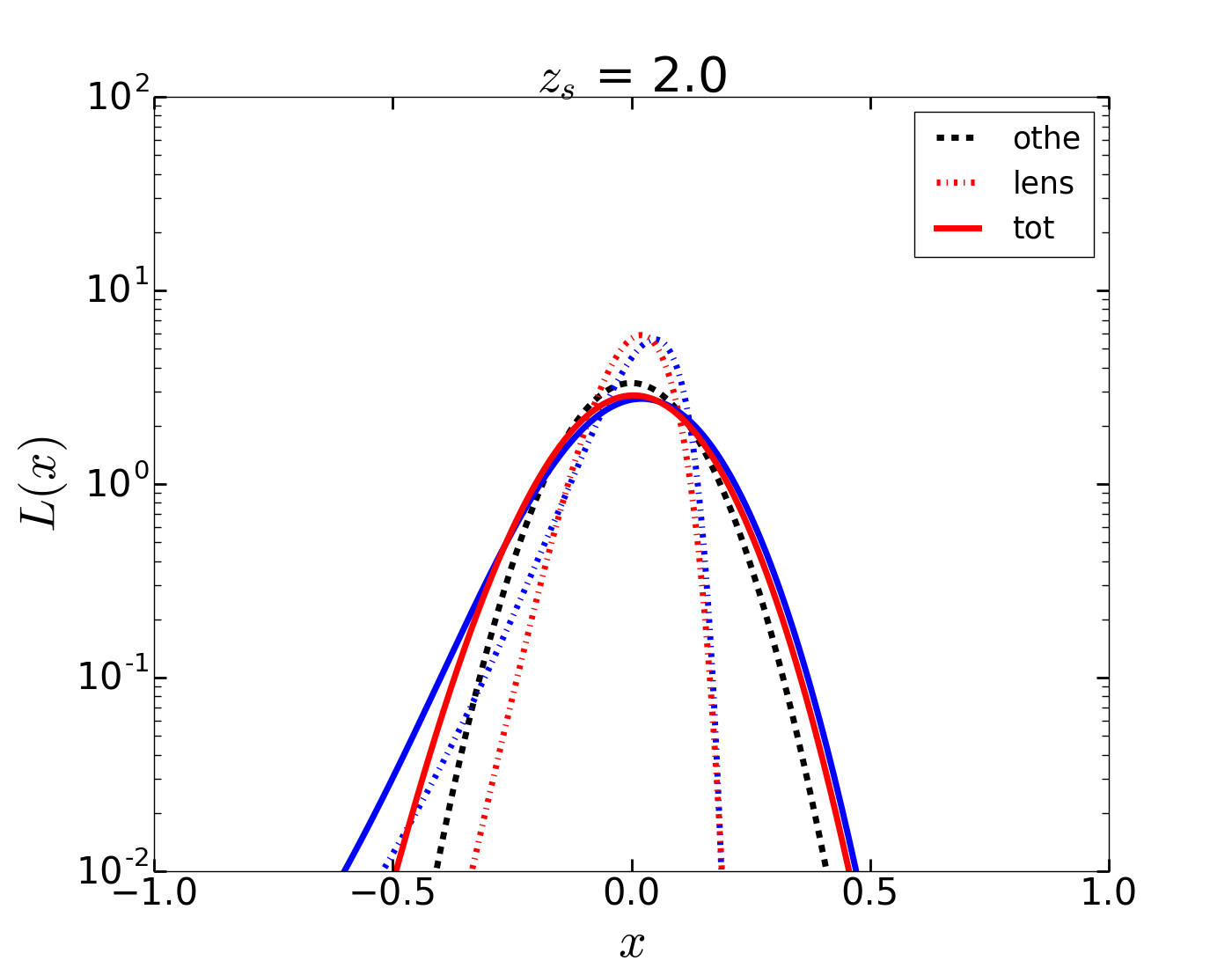}
  \end{center}
\end{minipage}
\end{tabular}
  \caption{\label{fig:PDF_x} PDFs of the apparent magnitude for various source redshifts. The dot-dashed, dashed (in black), and solid lines show the PDFs of the lensing effect $x_{\rm lens}$, other effects $x_{\rm othe}$, and the total scatter $x_{\rm tot}$, respectively. For comparison, we show our model, the lognormal model from \cite{2002ApJ...571..638T}, (in red) and the modified lognormal model from \cite{2011ApJ...742...15T} (in blue, only for $z_{s} = $ 1.0 and 2.0). The lensing effect becomes more dominant as the source redshift is higher.}
\end{figure}

\subsection{\label{mag_PDF}Magnitude PDF}

First, let us see the PDFs for the apparent magnitude. Fig.~\ref{fig:PDF_x} shows $L_{\rm lens}$, $L_{\rm othe}$, and $L_{\rm tot}$ at $z_{s} = $ 0.5, 1.0, 1.5, and 2.0 (in red). We can see that the PDF for the total scatter, $L_{\rm tot}$, more reflects the shape of $L_{\rm lens}$ at higher redshift. This is because the light rays from SNe Ia at higher redshift passes through the LSS for a longer time and gets more influenced by the weak-lensing effect, while the uncertainties other than the lensing effect, $x_{\rm othe}$, doesn't depend on redshift. In fact, we find that there is almost no difference between $L_{\rm othe}$ and $L_{\rm tot}$ because of the smallness of the lensing effect at low redshift: $z_{s} = 0.5$. We then, in the rest of paper, approximate $L_{\rm tot}$ by $L_{\rm othe}$ in the range of $z_s < 0.5$ in order to avoid the difficulty in the integral of Eq.~(\ref{eq:P_x_tot}).\footnote{$L_{\rm lens}$ is more peaky at lower redshift, which makes it harder to execute the convolution integral in Eq.~(\ref{eq:P_x_tot}). \label{foot_5}} Note that the panel of $z_{s} = 0.5$ also means that the value of $\sigma_{\rm othe}$ could be determined from observed data of SNe at low redshift.

While we use the simple lognormal function presented in \cite{2002ApJ...571..638T} as the convergence PDF, there are some other models. In~\cite{2006ApJ...645....1D}, they showed that the PDF of the projected surface mass density, which corresponds to the convergence in the thin mass-sheet case, calculated from $N$-body simulations is well described by a modified lognormal function. This modified lognormal distribution was tested against the simulated convergence along the line of sight (i.e., Eq.~(\ref{eq:K-z})) that was created by high-resolution ray-tracing simulations, and parameters in the model were calibrated so as to optimize the agreement with the simulations~\cite{2011ApJ...742...15T}. For comparison, we also show, in the panels of $z_{s} = $ 1.0 and 2.0, the magnitude PDFs  that are computed from the convergence PDF based on their modified lognormal function (in blue).\footnote{When calculating the PDFs for our model ($L_{\rm lens}$ etc.) in Fig.~\ref{fig:PDF_x}, we adopted the cosmological parameters from the WMAP 5-year result~\cite{2009ApJS..180..330K}, which was used as a fiducial cosmology in \cite{2011ApJ...742...15T}.\label{foot_6}} We can see that the difference between both types of PDFs becomes more apparent for smaller apparent magnitudes (i.e., higher magnification). This means that the simple lognormal function fails to describe the convergence PDF at high-magnification tails although we take account of the sample selection to exclude highly magnified samples (see Sec.~\ref{smal}).

\subsection{\label{fisher}Fisher matrix forecasts}

Next, we introduce the Fisher information matrix to study the cosmological constraints from SNe Ia. We consider a vector of a given data set ${\bf x} = (x_{1},\cdots,x_{N})$ and assume its probability distribution $f({\bf x}; {\bf \Theta})$ depends on a vector of model parameters ${\bf \Theta}=(\theta _{1},\cdots,\theta_{m})$. Regarding $f$ as the likelihood function, the Fisher information matrix is then defined as  
\begin{eqnarray}
	({\bf F})_{ij} \equiv -\left \langle \frac{\partial^{2} \ln f}{\partial \theta_{i} \partial \theta_{j}} \right \rangle,   \label{eq:fisher_mat}
\end{eqnarray} 
and its inverse ${\bf F}^{-1}$ gives the standard deviations for the errors on these parameters measured by the maximum likelihood estimate: $\sigma(\theta_{i})=({\bf F}^{-1})_{ii}$, where $\sigma(\theta_{i})$ is the standard deviation of the error on a parameter $\theta_{i}$~\cite[see][for a review]{1997ApJ...480...22T}. 

In our situation, the observed data vector $x_{i}$ is the set of apparent magnitudes $m_{{\rm obs},i}$ of SNe Ia at $z_{i}$ (see Eq.~(\ref{eq:dm})), 
\begin{eqnarray}
	m_{{\rm obs}, i} = m_{{\rm true}}(z_{i}) + \underbrace{x_{{\rm lens}, i} + x_{{\rm othe}, i}}_{x_{{\rm tot}, i}}  \label{eq:m_obs}
\end{eqnarray}
where
\begin{eqnarray}
	m_{{\rm true}}(z_{i}) = 5\log_{10}d_{\rm L}^{\rm FRW}(z_{i}) + {\rm const}.\label{eq:m_true}
\end{eqnarray}
Here $d_{\rm L}^{\rm FRW}[H_{0}, \Omega_{k}, \Omega_{m}, \Omega_{\Lambda}$, and $w]$ is the luminosity distance in a homogeneous FRW universe. Using the PDF, $L_{\rm tot}$, for $x_{\rm tot}$, the PDF for the $i$-th SN Ia is given by 
\begin{eqnarray}
	f_{i} = L_{\rm tot}\left[m_{{\rm obs}, i} - m_{{\rm true}}(z_{i})\right]. \label{eq:P_m_obs}
\end{eqnarray}
Assuming that there are no correlations between different SNe and between different types of errors, the probability distribution (or likelihood function) $f$ is    
\begin{eqnarray}
	f = \prod_{i=1}^{N} f_{i}.  \label{eq:f}
\end{eqnarray}
Finally, we can obtain the Fisher information matrix from Eqs.~(\ref{eq:fisher_mat}) and (\ref{eq:f}).

Unless otherwise noted, we fix the cosmological parameters to the values presented in Planck 2018 analysis~\cite{2018arXiv180706209P}, where the curvature density $\Omega_{k} = 0$, the matter density $\Omega_{m} = 0.316$, the baryon density $\Omega_{b} =  0.049$, the (cold) dark matter density $\Omega_{c} = 0.265$, the dark energy density $\Omega_{\Lambda} = 0.684$, the dark energy equation of state $w = -1$, and the total mass of neutrinos $\Sigma m_{\nu}=0.06\eV$. In the following subsections, we consider three model parameters: $\theta_{1}$ and $\theta_{2}$ are the cosmological parameters that we focus on; $\theta_{3} = \sigma_{\rm othe}$ is the uncertainty other than the lensing effect, which is always marginalized.

\subsubsection{\label{m_L}Density parameters: $\Omega_m$ and $\Omega_{\Lambda}$}

\begin{figure}[t]
\begin{tabular}{cc}
 \begin{minipage}{0.5\hsize}
  \begin{center}
   \includegraphics[width=80mm]{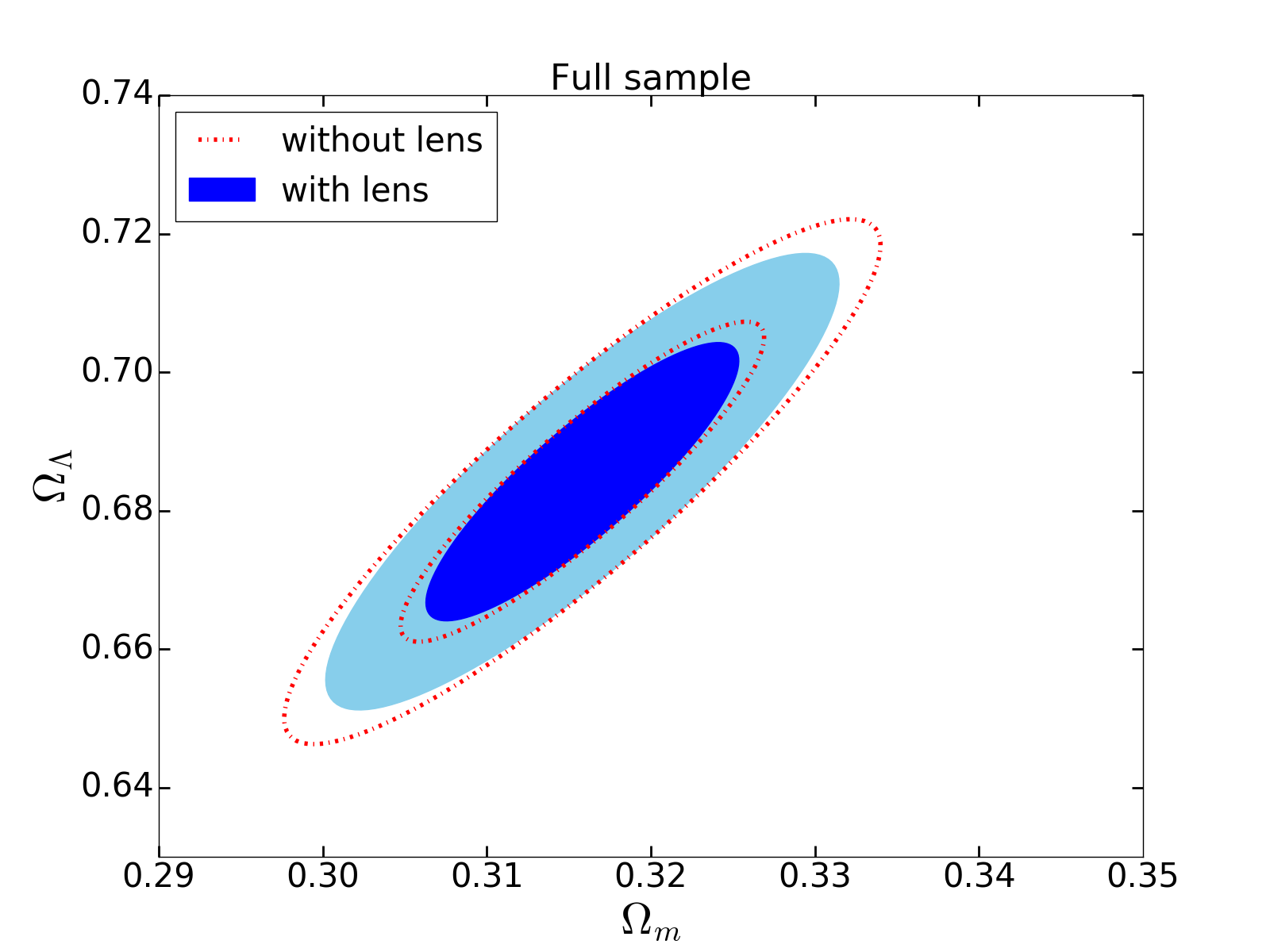}
  \end{center}
 \end{minipage}
 \begin{minipage}{0.5\hsize}
  \begin{center}
   \includegraphics[width=80mm]{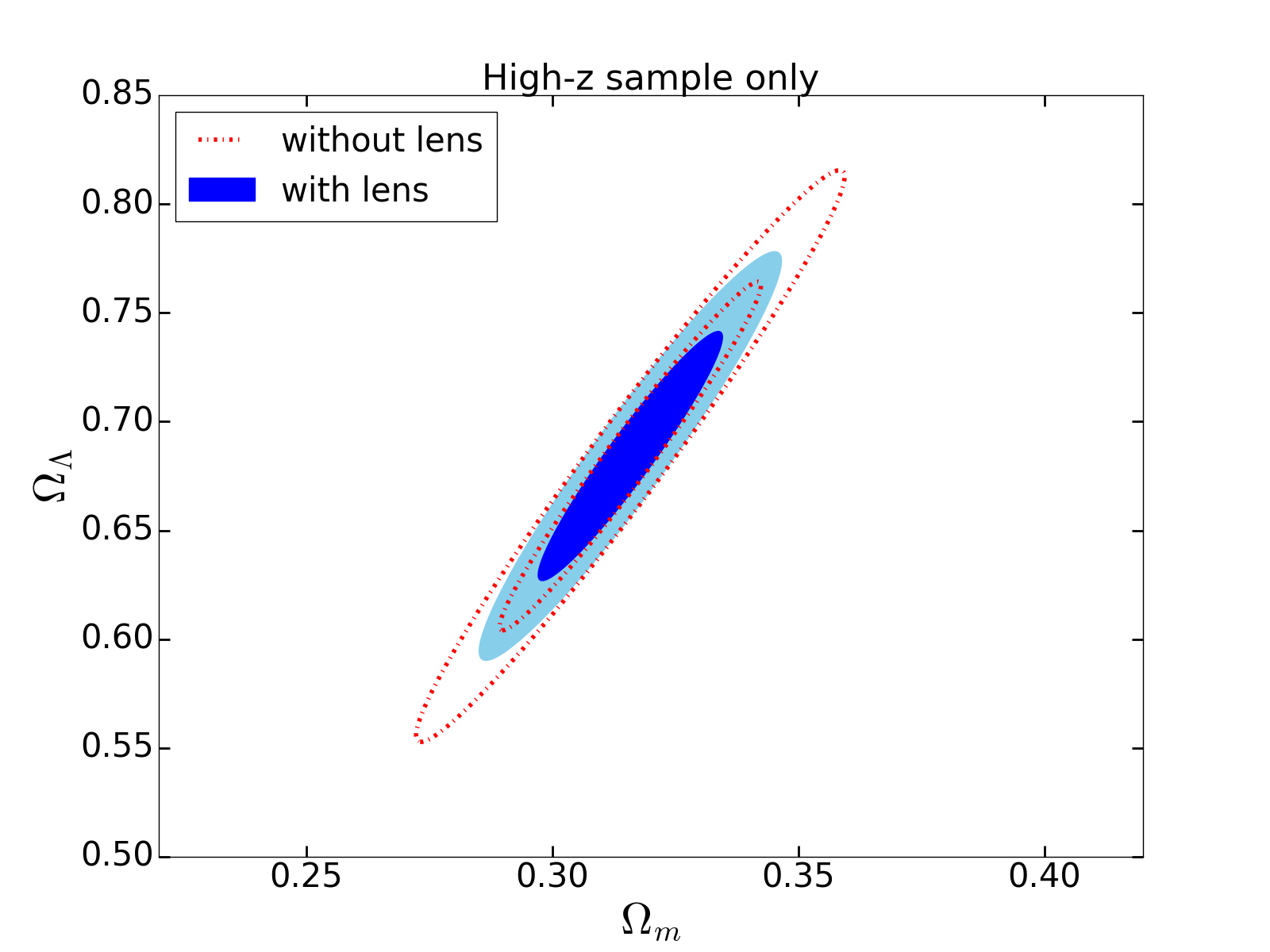}
  \end{center}
\end{minipage}
\end{tabular}
  \caption{\label{fig:m_vs_L} Comparison between the forecasts with (blue regions) and without (red dot-dashed lines) the weak-lensing effect, for the constraint on $\Omega_m$ and $\Omega_{\Lambda}$. {\it Left panel}: the $1\sigma\ (68\%)$ and $2\sigma\ (95\%)$ confidence contours from the {\it WFIRST} full sample. {\it Right panel}: for the high-$z$ sample ($z_{s} > 1.0$) only.}
\end{figure}

The left panel of Fig.~\ref{fig:m_vs_L} shows $1\sigma$ and $2\sigma$ contours of $(\theta_{1}, \theta_{2}) =  (\Omega_m, \Omega_{\Lambda})$ from the full SNe Ia sample in {\it WFIRST} (Table~\ref{table_1}). For comparison, we also show the result without considering the weak-lensing effect (red dot-dashed lines), which corresponds to using $L_{\rm othe}$ with the constant $\sigma_{\rm othe}^2 = \sigma_{{\rm othe}, f}^2 + \sigma_{\rm lens}^2(z_s)$ in Eq.~(\ref{eq:P_m_obs}) instead of $L_{\rm tot}$.\footnote{Note that while $\sigma_{\rm lens}$ is calculated by $\sigma_{\rm lens}^2(z_s) = (5/\ln 10)^2 \langle \kappa^2 (z_s)\rangle$ and dependent on the cosmological parameters, it is recognized as a constant at each redshift when implementing the Fisher analysis. \label{foot_7}} Here let us consider how the likelihood function $f$, i.e., the PDF $f_{i}$, depends on the cosmological parameters. In the case without the lensing effect, $f_{i}$ depends on $\Omega_m$ and $\Omega_{\Lambda}$ only through $m_{{\rm true}}$ (that is, $d_{\rm L}^{\rm FRW}$), which shifts the entire $f_{i}$ along the $m_{\rm obs}$-axis. In contrast, the PDF $f_{i}$ including the lensing effect depends on the cosmological parameters through not only $d_{\rm L}^{\rm FRW}$ but also $\kappa_{\rm min}$ and $\langle \kappa^2 \rangle$, which change the shape of $f_{i}$. This means that the likelihood with the lensing effect is more sensitive for the change of the cosmological parameters than the one without the lensing effect. We can actually see that the constraint with the lensing effect is slightly better than the one without the lensing effect.
 
On the other hand, the right panel of Fig.~\ref{fig:m_vs_L} shows the contours from only SNe Ia sample at higher redshift $z > 1$ (the high-$z$ sample). Compared with the result from the full sample, it is easier to see the improvement in the constraint due to considering the lensing effect although the constraint itself gets worse because of the lack of the low-$z$ sample. It reflects the fact that the contribution of the lensing effect is dominant at $z \gsim 1$ and the high-$z$ sample is only 36\% of the full sample. This result also say that if we can obtain SNe Ia at higher redshift in future, taking account of the lensing effect gets more important for the parameter constraints.

\subsubsection{\label{EoSoDE}Dark energy equation of state: $w$}

\begin{figure}[t]
\begin{tabular}{cc}
 \begin{minipage}{0.5\hsize}
  \begin{center}
   \includegraphics[width=80mm]{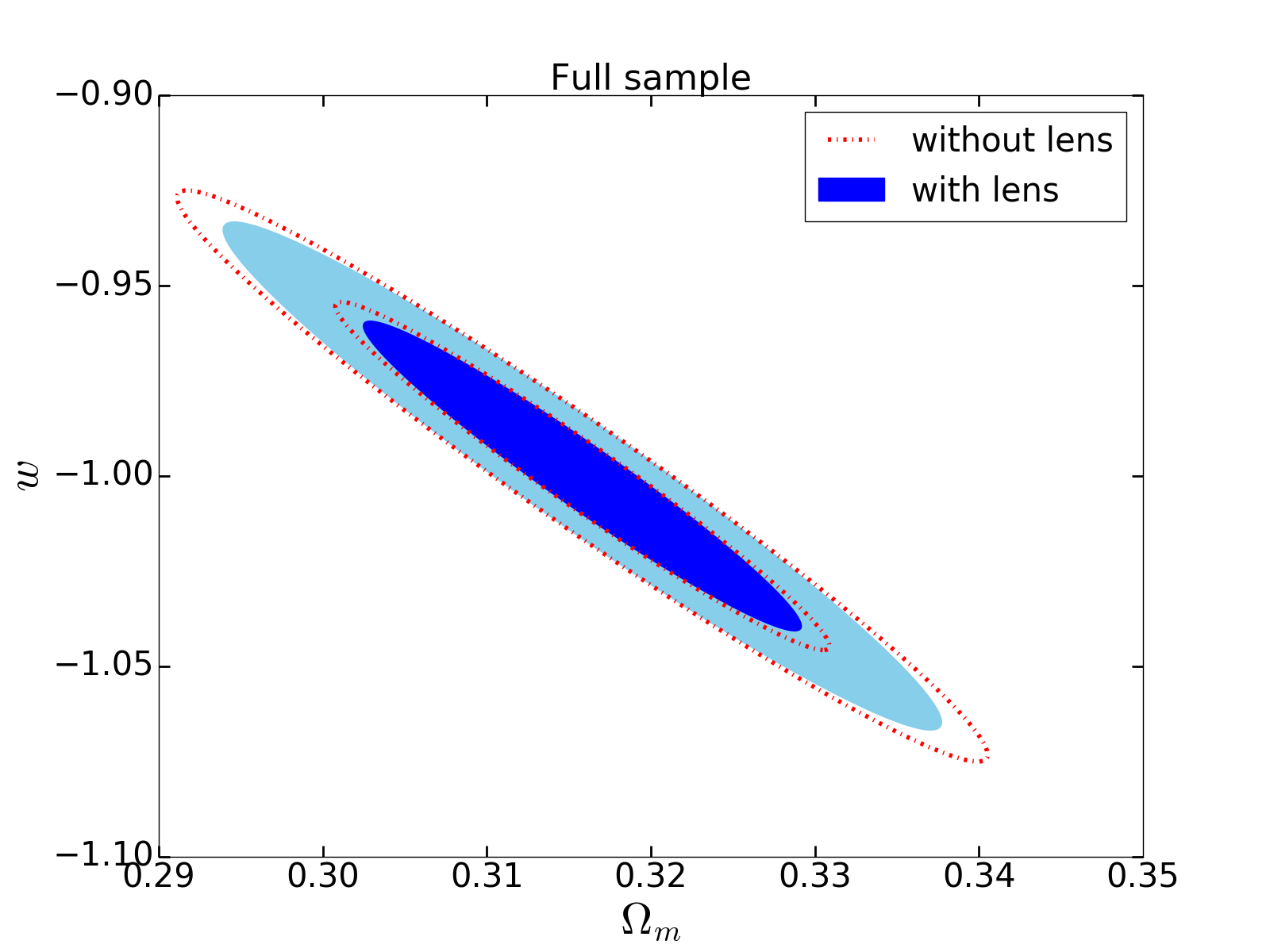}
  \end{center}
 \end{minipage}
 \begin{minipage}{0.5\hsize}
  \begin{center}
   \includegraphics[width=80mm]{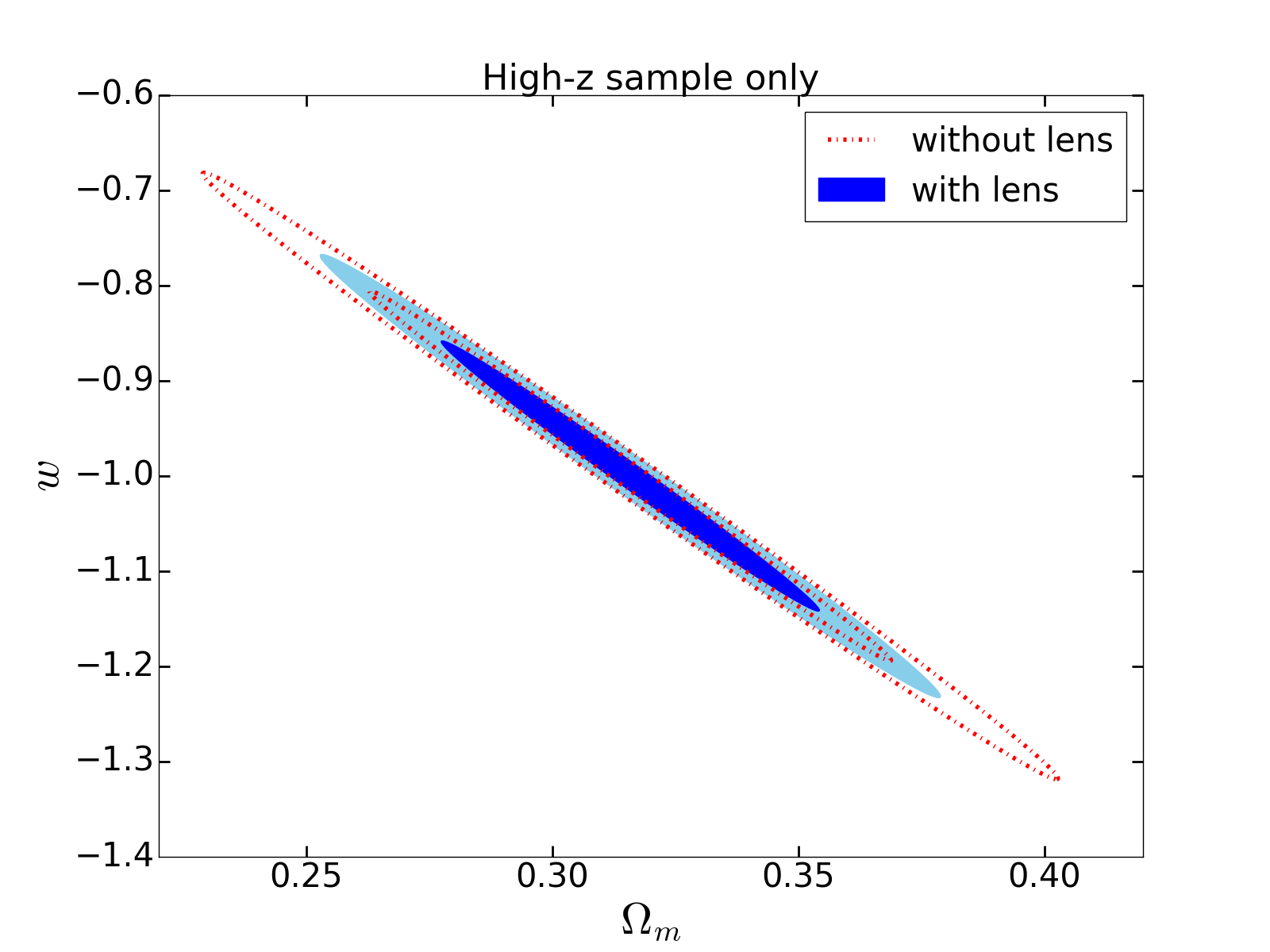}
  \end{center}
\end{minipage}
\end{tabular}
  \caption{\label{fig:m_vs_w} Comparison between the forecasts with (blue regions) and without (red dot-dashed lines) the weak-lensing effect, for the constraint on $\Omega_m$ and $w$. Each panel is described in the same manner as Fig.~\ref{fig:m_vs_L}.}
\end{figure}

Next, we turn our attention to the dark energy equation of state $w$. Fig.~\ref{fig:m_vs_w} shows the constraints on $(\theta_{1}, \theta_{2}) =  (\Omega_m, w)$, described in the same manner as Fig.~\ref{fig:m_vs_L}. Also in this case, we find that the constraint with the lensing effect is better than the one without the lensing effect, especially for the high-$z$ sample.

\subsubsection{\label{M_nu}Neutrino masses: $\Sigma m_{\nu}$}

\begin{figure}[t]
  \begin{center}
   \includegraphics[width=110mm]{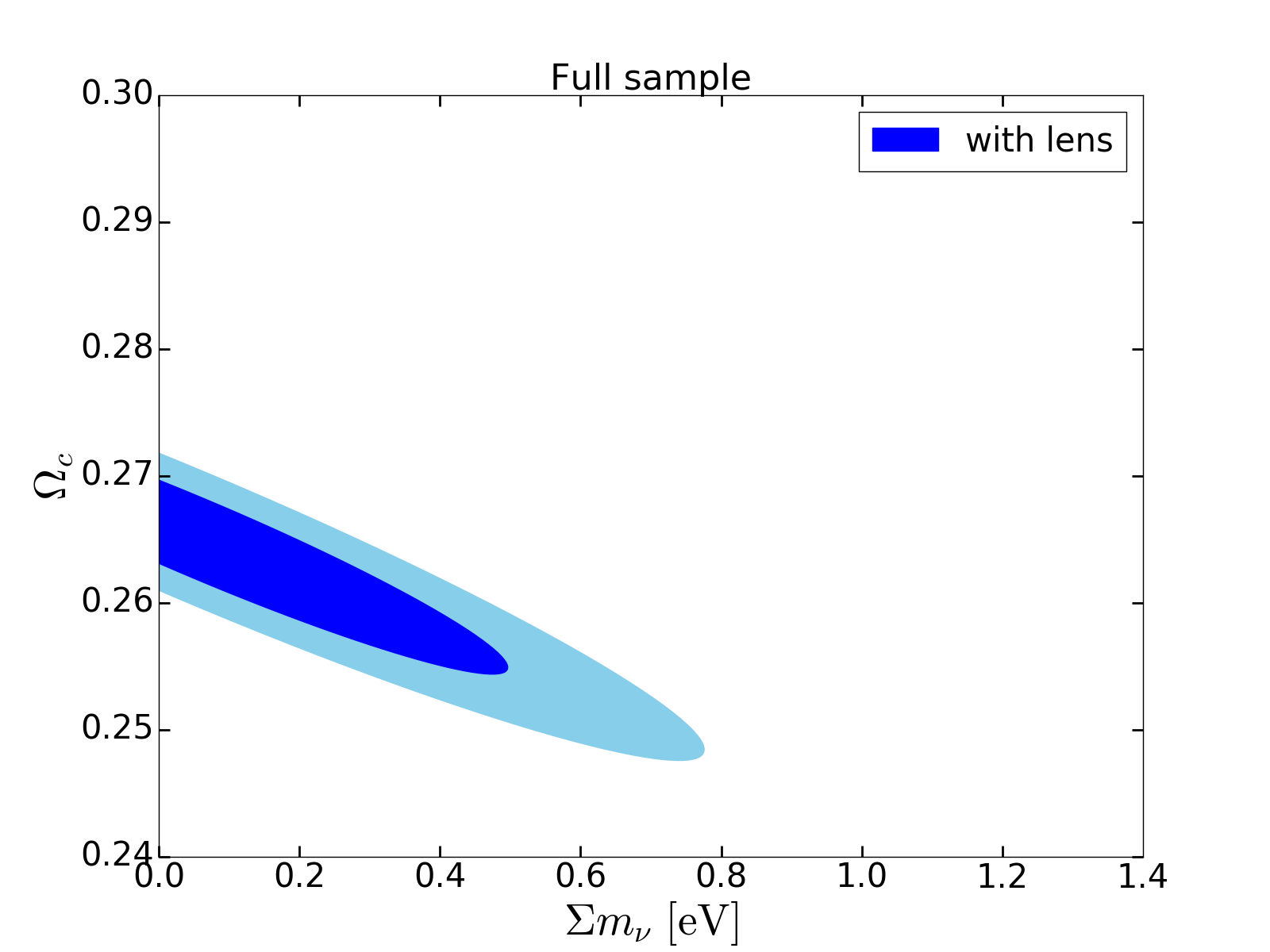}
  \end{center}
  \caption{\label{fig:nu_vs_c} The $1\sigma\ (68\%)$ and $2\sigma\ (95\%)$ confidence contours with the lensing effect (blue regions), for the constraint on $(\Sigma m_{\nu}, \Omega_c)$ from the {\it WFIRST} full sample.}
\end{figure}

Finally, let us consider of the possibility of constraining on the total mass of neutrinos $\Sigma m_{\nu}$ from SNe Ia. To do so, we focus on the following parameter set: $(\theta_{1}, \theta_{2}) =  (\Sigma m_{\nu}, \Omega_c)$. Neglecting the lensing effect, the PDF $f_{i}$ depends on the total mass of neutrinos $\Sigma m_{\nu}$ in $\Omega_m$ (only through $m_{{\rm true}})$: 
\begin{eqnarray}
	\Omega_{m} = \Omega_{c} + \Omega_{b} + \Omega_{\nu},\label{eq:Omega_m}
\end{eqnarray}
where $\Omega_{\nu} \propto \Sigma m_{\nu}$ (e.g., \cite{2006PhR...429..307L}). In the case with $(\theta_{1}, \theta_{2}) =  (\Omega_m, \Omega_{\Lambda})$, the terms with $\Omega_m$ and $\Omega_{\Lambda}$ in $d_{\rm L}^{\rm FRW}$ differently depend on the scale factor, $a$, and we can then constrain these parameters even not considering the lensing effect (see Sec.~\ref{m_L}). When it comes to $(\theta_{1}, \theta_{2}) =  (\Sigma m_{\nu}, \Omega_c)$, however, $\Sigma m_{\nu}$ (i.e., $\Omega_{\nu}$) and $\Omega_c$ equally contributes to $d_{\rm L}^{\rm FRW}$ through $\Omega_m$. It means that the these two parameters are completely degenerate, and we can not constrain them at all without the lensing effect.\footnote{In terms of the Fisher information matrix, the complete degeneracy corresponds to $| {\bf F} | = 0$. \label{foot_8}}

On the other hand, as we mentioned in Sec.~\ref{m_L}, the PDF $f_{i}$ including the lensing effect depends also on $\langle \kappa^2 \rangle$. Massive neutrinos suppress the growth of the linear and non-linear matter power spectrum on small scales, and the total mass of neutrinos $\Sigma m_{\nu}$ consequently changes the variance $\langle \kappa^2 \rangle$, independently of $\Omega_c$ (for details, see Sec.~3.1 in \cite[][]{2016ApJ...828..112H}). Thus it is expected that taking into account of the weak-lensing convergence, we can resolve the degeneracy between $\Sigma m_{\nu}$ and $\Omega_c$, and constraint $\Sigma m_{\nu}$ from SNe Ia. The left panel of Fig.~\ref{fig:nu_vs_c} shows the constraints of $(\theta_{1}, \theta_{2}) =  (\Sigma m_{\nu}, \Omega_c)$ from the full {\it WFIRST} sample. We can see that the degeneracy is actually resolved and the expected upper bound is $\Sigma m_{\nu} < 0.78\eV$ (95\% CL).

\section{\label{conc}Discussion and Conclusions}

In this paper, we have investigated how the cosmological constraints from SNe Ia are improved by including the weak-lensing convergence. Forecasting the parameter constraints for the {\it WFIRST} samples, we showed that because of the LSS information from the weak-lensing effect, the constraints on $ (\Omega_m, \Omega_{\Lambda})$ and $(\Omega_m, w)$ are slightly improved. In addition, limiting to only SNe Ia sample at higher redshift $z > 1$, we can more clearly see the improvement. This result reflects the fact that the contribution of the lensing effect to the PDF of the magnitude scatter is dominant at $z \gsim 1$, which we showed in Fig.~\ref{fig:PDF_x}. Furthermore, we focused on $\Sigma m_{\nu}$ and $\Omega_{c}$, and found that the degeneracy between these two parameters can be resolved by considering the lensing effect and the upper bound on the total mass of neutrinos expected from the {\it WFIRST} survey is $\Sigma m_{\nu} < 0.78\eV$ (95\% CL).

It is known that there are two possibilities for the mass hierarchy of neutrinos: the normal hierarchy $\Sigma m_{\nu} \gsim 0.06\eV$ and the inverted hierarchy $\Sigma m_{\nu} \gsim 0.1\eV$ (e.g., \cite{2006PhR...429..307L}). Setting an upper bound $\Sigma m_{\nu} < 0.1\eV$ therefore leads to ruling out the inverted hierarchy, which would have significant impact on particle physics. In that sense, the expected constraint from SNe Ia in {\it WFIRST} is insufficient to set such a upper bound. However, it can be improved in the future. As a trial, if we perform the forecast in Sec.~\ref{m_L} with $\sigma_{{\rm othe},f}=0.06$ (half of the assumption in {\it WFIRST}), the upper bound $\Sigma m_{\nu} < 0.34\eV$ (95\% CL) for $(\theta_{1}, \theta_{2}) =  (\Sigma m_{\nu}, \Omega_c)$ is obtained ($\theta_{3} = \sigma_{\rm othe}$ is marginalized again). It is actually expected that the uncertainty $\sigma_{{\rm othe},f}$ will be reduced by correcting a large amount of SNe Ia data at low redshift, where the lensing effect can be neglected. For instance, while SNe Ia at $z < 0.7$ expected from LSST (main) is not effective for constraining neutrino masses, we can utilize the large number of SNe Ia $O(10^4)$ for reducing $\sigma_{\rm othe}$.

Here we comment on the convergence PDF. As we mentioned in Sec.~\ref{mag_PDF}, the simple lognormal function that we used in this work cannot properly describe high-magnification tails of the modified lognormal function that is calibrated against a ray-shooting simulation~\cite{2006ApJ...645....1D, 2011ApJ...742...15T}. However, this modified lognormal model is not given as a function of the cosmological parameters and thus we need some efforts to apply it to the parameter forecast. In addition, the sample selection model for SNe Ia to avoid highly magnified samples has some assumptions and uncertainties (for details, see Sec.~4 in \cite[][]{2016ApJ...828..112H}). Therefore, it is still worth calibrating our sample selection method by ray-tracing simulations and finding out how the convergence PDF is improved.

In any case, combining the weak-lensing effect, SNe Ia still have the potential to be an independent and interesting probe of the LSS, especially neutrino masses. We have been focusing on SNe Ia as {\it standard candles} so far, however, the idea of extracting the information of LSS through the weak-lensing convergence can be also applied to {\it standard sirens}. In particular, binary neutron stars (BNSs) draw attention as potential candidates for standard siren because its electromagnetic (EM) counterpart can be detected at the same time, which allows us to determine their location of the sky and redshift. In fact, recently gravitational waves from a BNS merger and its EM counterpart were, for the first time, observed by Advanced LIGO~\cite{2017PhRvL.119p1101A}. Moreover, the Einstein Gravitational-Wave Telescope (ET)\footnote{\url{http://www.et-gw.eu}}~\cite{ET2011}, which is a next-generation gravitational wave detector, can observe the GW signals from BNSs up to $z \simeq 2$. For BNS mergers in the redshift range $1 < z < 2$, the lensing effect is dominant, and thus utilizing BNSs allow us to extract the information of LLS as well as SNe Ia, although we still need to pay attention to the distance uncertainty other than the lensing effect. In \cite{2010CQGra..27u5006S}, by extrapolating the rate of BNS inspirals expected in advanced detectors and considering that GRBs, as EM counterparts, are beamed with a beaming angle, it is conservatively assumed that the ET would observe about $10^3$ BNS mergers that have EM counterparts over a 5 year period. Supposing that all sources were
distributed uniformly in $0 < z < 2$ and $\sigma_{{\rm othe},f}=0.02$ (which corresponds to a distance accuracy of $\delta d/d \sim 1\%$), we obtain the upper bounds on the total mass of neutrinos from our forecast $(\theta_{1}, \theta_{2}) =  (\Sigma m_{\nu}, \Omega_c)$: $\Sigma m_{\nu} < 0.26\eV$ (95\% CL) for $10^3$ BNS mergers and positively, $\Sigma m_{\nu} < 0.13\eV$ (95\% CL) for $10^4$ BNS mergers ($\theta_{3} = \sigma_{\rm othe}$ is marginalized again). This result suggests that the ET could allow us to reach the criterion for the inverted hierarchy $\Sigma m_{\nu} \simeq 0.1\eV$ if we can achieve a distance accuracy of $1\%$, and BNS mergers can be a more powerful probe of the LSS than SNe Ia.

\acknowledgments

We would like to appreciate R. Takahashi for providing the data from his ray-shooting simulation. We also thank M. Takada for useful discussions. RH is supported by World Premier International Research Center Initiative (WPI), MEXT, Japan and JSPS Research Fellowships for Young Scientists (No.~19J00513). TF is supported by Grant-in-Aids for Scientific Research from JSPS (No.~17K05453 and No.~18H04357).

\bibliographystyle{JHEP.bst}
\bibliography{ms.bib}

\end{document}